\documentclass{pjab}
\usepackage[T1]{fontenc}
\usepackage{lmodern}
\usepackage{textcomp}
\usepackage{amsmath}
\usepackage[dvipdfmx]{graphicx}
\usepackage{array}
\usepackage[superscript]{cite}

\usepackage{color}

\begin{document}
\pjabcategory{Original Article}
\title[Eleven-year and multidecadal solar cycles in a Dome Fuji ice core]
      {Eleven-year, 22-year and $\sim$90-year solar cycles 
discovered \\ in nitrate concentrations in a Dome Fuji (Antarctica) ice core
}
\authorlist{%
 \Cauthorentry{Yuko Motizuki}{labelA,labelB}
 \authorentry{Yoichi Nakai}{labelA}
 \authorentry{Kazuya Takahashi}{labelA}
 \authorentry{Takashi Imamura}{labelC,labelD}
  \authorentry{Hideaki Motoyama}{labelE,labelF}
}
\affiliate[labelA]{RIKEN Nishina Center, Wako, Japan}
\affiliate[labelB]{Graduate School of Science and Engineering, Saitama University, Saitama, Japan}
\affiliate[labelC]{National Institute for Environmental Studies, Tsukuba, Japan}
\affiliate[labelD]{Present address: Tokyo Metropolitan Research Institute for Environmental Protection, Tokyo, Japan}
\affiliate[labelE]{National Institute of Polar Research, Tachikawa, Tokyo, Japan}
\affiliate[labelF]{The Graduate University for Advanced Studies, SOKENDAI, Tachikawa, Tokyo, Japan}
\Correspondence{Y. Motizuki, RIKEN Nishina Center, Hirosawa 2-1, Wako, Saitama 351-0198, Japan (motizuki@riken.jp)}
\abstract{
Ice cores are known to yield information about astronomical phenomena as well as information about past climate. 
We report time series analyses of annually resolved nitrate variations in an ice core, drilled at the Dome Fuji station in East Antarctica, corresponding to the period from CE 1610 to 1904.
Our analyses revealed clear evidence of  $\sim$11,  $\sim$22, and $\sim$90 year periodicities, 
comparable to the respective periodicities of the well-known Schwabe, Hale, and Gleissberg solar cycles. 
Our results show for the first time that nitrate concentrations in an ice core can be used as a proxy for past solar activity on decadal to multidecadal time scales. 
Furthermore, 11-year and 22-year periodicities were detected in nitrate variations even during the Maunder Minimum (1645--1715), when sunspots were almost absent.
This discovery may support cyclic behavior of the solar dynamo during the grand solar minimum. \\
}
\maketitle

\section{Introduction} \label{sec:intro}

Traditionally, cosmogenic nuclides, $^{14}$C in tree rings
\cite{Bard1997}$^-$\cite{Heaton2021}
and $^{10}$Be in ice cores
\cite{Beer1990}$^-$\cite{Horiuchi2008}
have been used to investigate past solar activity cycles.
Brehm et al.\cite{Brehm2021} found an 11-year solar cycle in annually resolved $^{14}$C tree sample measurements covering the last millennium, but 
they could not identify an 11-year cycle during the Maunder Minimum (1645--1715),\cite{Eddy1976}  when sunspots were almost absent, because of weather-induced noise. 
Annually resolved $^{10}$Be data from 
a Greenland ice core record an 11-year solar cycle during the last 600 years (1400--2000),\cite{Berggren2009} even during the Maunder Minimum, despite the presence of weather-induced noise. 
These tree-ring $^{14}$C records and ice core $^{10}$Be records do not agree well, 
neither with respect to observations of 11-year oscillations in the grand solar minimum
nor with respect to oscillations on time scales of 1,000 years and longer.\cite{Heaton2021}
It is thus important to seek another potential proxy for solar activity. 
In the present study, we investigated the nitrate ion (NO$_3^-$) concentration in an Antarctic Dome Fuji ice core as a potential new proxy.

\begin{figure*}[h!]
\begin{center}
\includegraphics[width=16.5cm,pagebox=cropbox,clip]{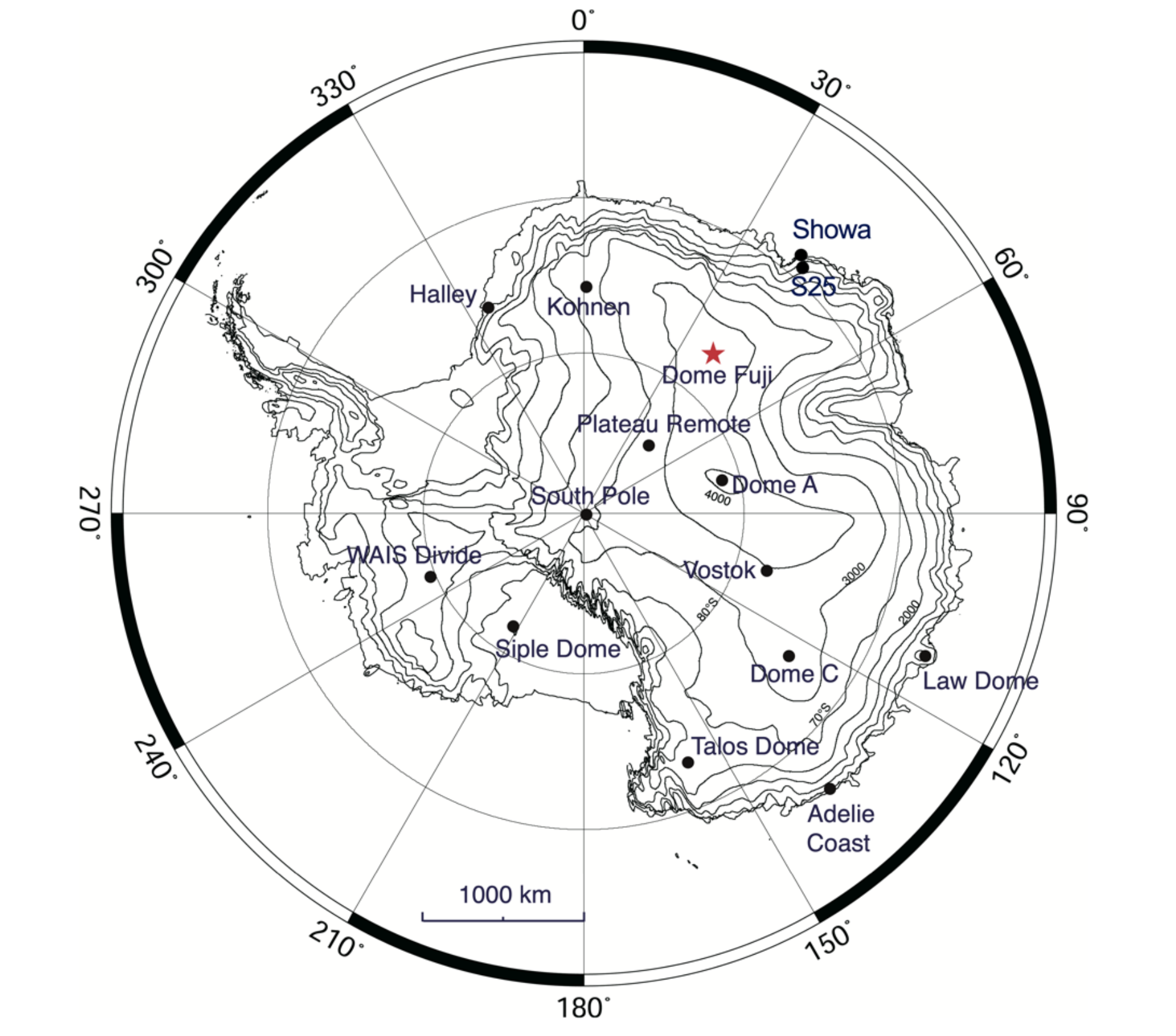}
\caption{
Map of Antarctica showing the locations of Dome Fuji station (red star) and other research stations and ice-core drilling sites, including the S25 core site mentioned in this study.
\label{fig:Antarctica}}
\end{center}
\end{figure*}

It has been known that continental and anthropologic effect in NO$_3^-$ concentrations in Antarctic snow and ice are not significant.
\cite{LegrandDelmas1986}
NO$_3^-$  data obtained from Antarctic ice cores then suggested there might be a relationship between NO$_3^-$ and solar activity.
\cite{ZellerParker1981}$^-$\cite{Traversi2012} 
For example, Watanabe et al.\cite{Watanabe1999} 
demonstrated the existence of an 11-year cycle in NO$_3^-$ concentrations in a firn core covering a 60-year period (1920--1980) collected from the S25 site near Showa station in coastal East Antarctica  (Fig.~\ref{fig:Antarctica}). 
Traversi et al. (2012)\cite{Traversi2012} studied NO$_3^-$ concentrations in 
an ice core recovered from Talos Dome (Fig.~\ref{fig:Antarctica}). 
They noted that meteorological noise on an interannual scale made it impossible to resolve individual 11-year solar cycles and reported a weak statistical correlation (r = 0.31) between variations in solar activity-related 
cosmic ray intensities and the NO$_3^-$ concentrations in a segment 
corresponding to the ``pre-industrial'' period (1713--1913). 
With deeper ice core analyses, they concluded that NO$_3^-$ concentrations in the Talos Dome ice core were a potential new proxy for solar activity on centennial to millennial time scales. 
   
Extraction of the 11-year solar cycle from NO$_3^-$ concentration variations in ice cores is generally considered to be difficult, 
mainly because the process of NO$_3^-$ subsidence to the ground and subsequent preservation in ice cores is affected by atmospheric dynamics in the troposphere and recycling process from central Antarctica to the coast, 
even if the NO$_3^-$ budget in the stratosphere follows the solar modulations.
\cite{Wolff2008}
Furthermore, the NO$_3^-$ may have several terrestrial origins: 
For example, NO$_3^-$ is generated by lightning and biomass burning. 
\cite{LegrandDelmas1986}
\cite{LegrandKirchner1990}
\cite{Wolff1995}
As well, sporadically enhanced deposition of  NO$_3^-$ in sea salt and terrestrial aerosols
\cite{Wolff2008}
can affect NO$_3^-$ concentrations in ice cores and, due to its chemical properties, NO$_3^-$ evaporates easily and can be displaced by  other anions, 
such as sulfate (SO$_4^{2-}$) that form strong acids (see Sect.~\ref{sec:solarsig}).
Nevertheless, despite these difficulties, NO$_3^-$ constitutes one of the main anionic species in ice cores, so much effort has been put into understanding its origins and concentration profiles.
  
\begin{figure*}[ht!]
\begin{center}
\includegraphics[width=17.5cm,pagebox=cropbox,clip]{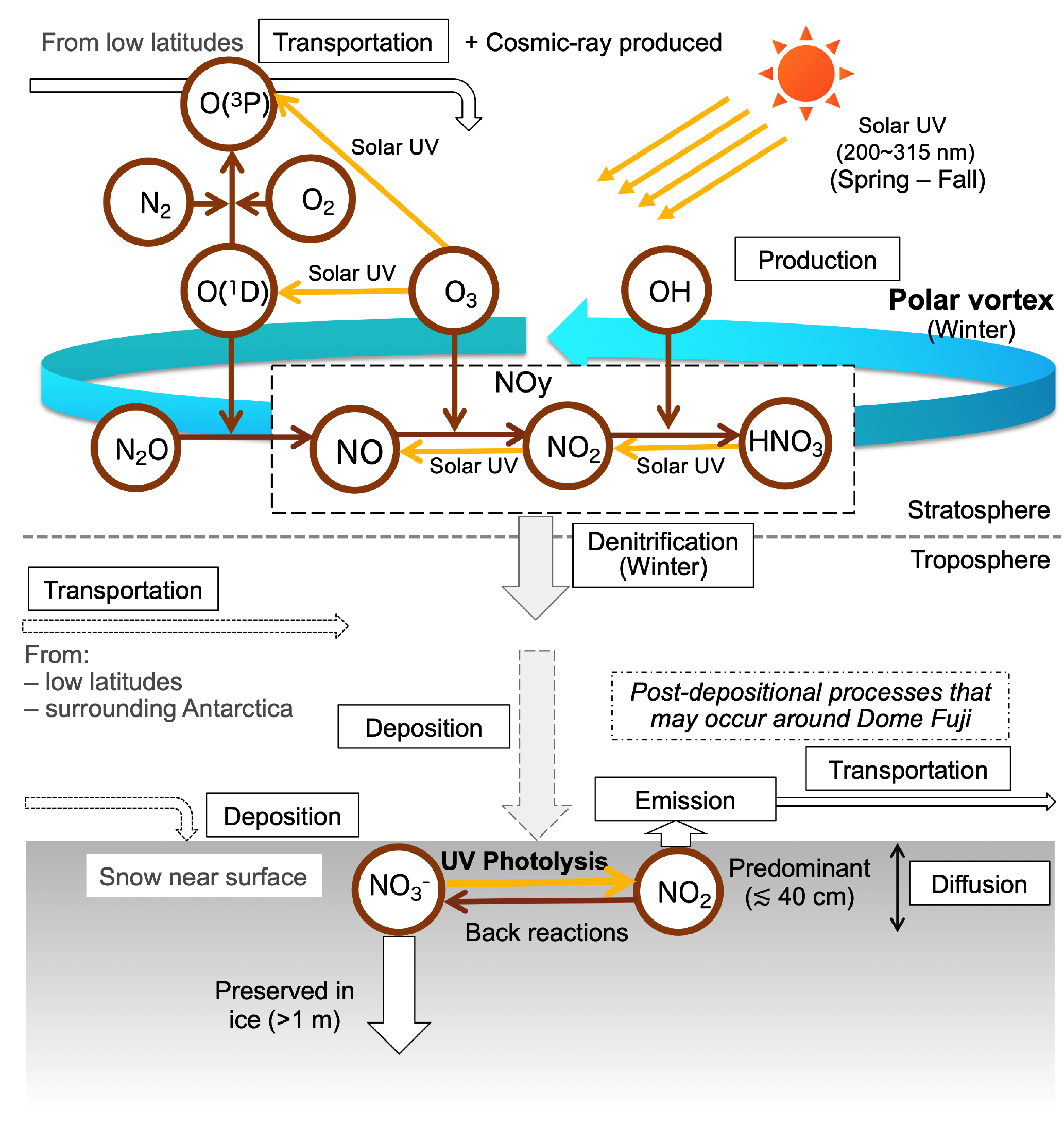}
\caption{
Schematic diagram of the stratospheric production of nitrogen oxides induced by solar UV radiation (wavelength mainly 200--315 nm) and post-depositional processes that likely occur around Dome Fuji station, where the snow accumulation rate is low. Only the principal chemical reaction chains are depicted; in particular, reaction channels to N$_2$O$_5$ are omitted. Denitrification occurs within the polar vortex, which develops in winter. 
Note that processes occurring at both microscopic and macroscopic scales are shown on this diagram.
\label{fig:reaction}}
\end{center}
\end{figure*}

Solar UV radiation (wavelength in 200--315 nm) is absorbed into the stratosphere and induces the production of reactive nitrogen, designated by NO$_{\rm y}$ (represented simply by NO, NO$_2$, and HNO$_3$ in this work), from atmospheric N$_2$O (Fig.~\ref{fig:reaction}).
\cite{BrasseurSolomon2005}\cite{VittJackman1996}
As discussed by Vitt and Jackman,\cite{VittJackman1996} 
the oxidation of N$_2$O by solar UV dominates the global NO$_{\rm y}$ source, but galactic cosmic rays (GCRs), which descend along geomagnetic field lines to the polar regions, are also significant at polar latitudes.
The effects of in situ production of NO$_{\rm y}$ by solar UV radiation, and their transport from lower ($<$50$^{\circ}$) latitudes, 
on the amount of NO$_{\rm y}$ in the polar stratosphere are likely to be larger than the effect of NO$_{\rm y}$ production by GCRs\cite{VittJackman1996} (see Sect.~\ref{sec:solarsig}).
 
The NO$_{\rm y}$ in the polar stratosphere is then deposited in precipitation down to the troposphere (a process called denitrification) 
within the polar vortex that develops in Antarctic winters (Fig.~\ref{fig:reaction}), which is where polar stratospheric clouds (PSCs) form.
The NO$_{\rm y}$, which constitute the aerosols in PSCs, are transported downwards into the troposphere by gravitational sedimentation,
and are then precipitated onto the surface snow.\cite{Lambert2016}\cite{Grooss2014}
NO$_{\rm y}$ thus accumulates in ice cores and, when ice core samples are melted, the NO$_{\rm y}$ occurs as aqueous NO$_3^-$ ions.
Since no solar UV radiation reaches Antarctica in wintertime, no back reactions (Fig.~\ref{fig:reaction}) operate, so the stratospheric NO$_{\rm y}$ in the Antarctic winter is composed mainly of HNO$_3$. 

Inclusion of the NO$_{\rm y}$ budget of stratospheric origin in ice cores 
is site-specific, and depends on the location of the drilling site: NO$_3^-$ in Antarctic ice cores drilled at coastal sites is dominated by the components from the troposphere\cite{LegrandDelmas1986}, whereas cores obtained from inland sites, in particular where effectively affected by the denitrification process,
are likely to contain some stratospheric NO$_3^-$ components, as in the case of Dome Fuji cores mentioned below. 
As a result, some ice cores from inland Antarctica are likely to contain a greater proportion of stratospheric NO$_{\rm y}$ than some cores from coastal Antarctica.
Solar UV radiation affects both the NO$_{\rm y}$ production in the stratosphere (200--315 nm) and the loss of NO$_3^-$ on the surface (300--340 nm).
In particular, there are post-depositional processes\cite{Roethlisberger2002} that also affect specific sites and greatly reduce the NO$_3^-$ concentrations in ice cores collected from low-accumulation sites:
the most relevant post-depositional process is 
photolysis in surface snow, caused by solar UV (300--340 nm, mainly UV-A) reaching the ground.
\cite{Frey2009}$^{-}$\cite{Akers2022}
 
Dome Fuji station (77$^{\circ}$ 19' 01'' S, 39$^{\circ}$ 42' 12'' E)
is located at an inland Antarctic site (Fig.~\ref{fig:Antarctica}),
on the summit of a mountain in Dronning Maud Land (elevation 3,810 m a.s.l.).
The mean annual temperature of snow at Dome Fuji from 1995 to 2006, measured at 10 m depth, was $-$57.3 $^{\circ}$C, and the mean rate of snow accumulation was 27.3 mm water-equivalent yr$^{-1}$. 
\cite{Kameda2008}
Snow and ice at Dome Fuji station may contain a relatively large fraction of stratospheric chemical components with respect to tropospheric components 
compared with some other sites in Antarctica, based on the following experimental evidence:

(1) Radioactive tritium (T) from fallout from nuclear bomb tests conducted in the 1960s, which is 
a key tracer of stratospheric subsidence, is found in ice cores. 
In fact, T concentrations in snow samples collected from around the Dome Fuji site were the highest measured in any of the samples collected from snow pits at 16 Antarctic sites 
(Fourr$\acute{\rm e}$ et al., 2006\cite{Fourre2006}), 
including Dome C, the Halley Research Station, the South Pole, Talos Dome, and Vostok (Fig.~\ref{fig:Antarctica}). 
Note, however, that Fourr$\acute{\rm e}$ et al. 
(their Table~1) incorrectly reported the highest T value (4200 Tritium Units, where 1 TU indicates a T/$^1$H ratio of 10$^{-18}$) 
in a sample obtained around Dome Fuji station as being from Dome C 
(at the time the paper, Kamiyama et al. (1989)\cite{Kamiyama1989}, cited by Fourr$\acute{\rm e}$ et al. as the reference for the highest 4200 TU value, was published, Dome Fuji station had yet to be constructed; but a site neighboring the present Dome Fuji station was referred to as  ``Dome Camp'' with the abbreviation ``DC'', which Fourr$\acute{\rm e}$ et al. misinterpreted as ``Dome C''.)

(2) Both wet and dry deposition containing tropospheric components also contain sea salt. However, the ratios of the averaged ionic concentrations in snow and ice samples from Dome Fuji are inconsistent with those that would be expected if most of the ions were of tropospheric origin. 
This result also suggests that the ionic components of samples from Dome Fuji may contain a higher proportion of stratospheric origin.
\cite{Kamiyama1989}$^{-}$\cite{Motizuki2017}
 
The denitrification process affecting Dome Fuji, as portrayed above, is supported by a year-round observation conducted in 1997--1998 by the 38th Japanese Antarctic Research Expedition.
\cite{Motoyama2005}
The NO$_3^-$ concentration in fresh snowfall observed at Dome Fuji started to increase drastically from late winter (July), giving a prominent peak in early spring (August), and the NO$_3^-$ enhancement continued until the end of spring (October).  
The NO$_3^-$ concentrations observed in the spring (Aug--Oct, 1997) were five times larger than those observed in the fall (Feb--Apr, 1997) and no summer peak was observed; these features are distinctly different from those observed at other Antarctic sites.
\cite{Wolff2008}\cite{Traversi2017}\cite{Legrand2017}
Also, the increased proportions of the NO$_3^-$ ions in equivalent concentrations from July to October in 1997 were accompanied by exactly the same increased amount of H$^+$ as counter cations. 
This indicates that the molecules in the depositional process were in the form of gaseous HNO$_3$.
As mentioned above, these observations were consistent with a picture that the NO$_3^-$ concentrations in Dome Fuji ice cores were affected by a denitrification process associated with the formation of the polar vortex and the PSCs, with a precipitation lag of about 1 to 2 months (see Fig.~\ref{fig:reaction}).
A more detailed study of the seasonal variations in ionic compositions observed at Dome Fuji will be presented elsewhere.
  
The explanation for the relatively high apparent stratospheric contribution to the Dome Fuji ice cores could be 
that the elevation of the site is high
(3,810 m a.s.l) and that the specific location may be in an area that tends to be affected by the denitrification process.
Although a detailed mechanism to explain the relatively higher stratospheric contribution around the Dome Fuji area compared with other areas of Antarctica should be investigated further, a Dome Fuji core appears to have reasonable potential for studying stratospheric NO$_{\rm y}$.
   
Because the snow accumulation rate at Dome Fuji is low, HNO$_3$
(including NO$_3^-$; see Fig.~\ref{fig:reaction}) precipitated in snow may undergo photolysis by solar UV radiation, 
as mentioned above,
leading to emission from or diffusion within the surface snow (Fig.~\ref{fig:reaction}).
In fact, NO$_3^-$ concentrations in the top $\sim$40 cm of snow at the Dome Fuji site were observed to be decreased very rapidly with depth; the concentration at 1 m snow depth can be as low as about one-tenth of the concentration at the snow surface (see fig. 6 of Watanabe et al.\cite{Watanabe2003}). 
This rapid decrease with depth implies that the post-depositional loss of NO$_3^-$ at Dome Fuji may be controlled predominantly by NO$_3^-$ photolysis, as reported for 
Kohnen station located in the same Dronning Maud Land area (Fig.~\ref{fig:Antarctica}).\cite{Winton2020}
The photolysis effect on the surface snow may greatly affect the NO$_3^-$ concentration profile recorded in Dome Fuji ice cores. 
This topic is also examined in Sect.~\ref{sec:solarsig}.

\section{The DF01 ice core: Drilling, analyses, and dating}\label{sec:DF01}

We investigated the uppermost part of a 122-m-long firn core (DF01) drilled in 2001 at Dome Fuji station. 
Core DF01 was obtained from the same hole as a deep ice core (DF2; 3,035.22 m) drilled over the course of the 7 years from 2001--2007
\cite{Motoyama2007}\cite{Motoyama2021} 
and extending back to $\sim$720,000 years ago\cite{Kawamura2017}.
The upper, younger part of the DF01 firn core was very fragile, and some portion of the top 7.7 m of the core was lost during the drilling. 
We performed continuous, annually resolved measurements of ions in the DF01 ice core segment from 7.7 to 85.5 m depth at RIKEN Nishina Center. 
 
The concentrations of anions 
($\rm{SO}_{4}^{\, \, 2-}$, $\rm{Cl}^-$, $\rm{NO}_{3}^-$, $\rm{F}^-$, $\rm{CH_{3}COO}^-$, 
$\rm{HCOO}^-$, $\rm{NO}_{2}^-$, $\rm{C}_{2}{O}_{4}^{2-}$, $\rm{PO}_{4}^{3-}$, and $\rm{CH_{3}SO}_{3}^-$)
and cations 
($\rm{Na}^+$, $\rm{K}^+$, $\rm{Mg}^{2+}$, $\rm{Ca}^{2+}$, and $\rm{NH}_{4}^+$)
in the DF01 ice core were analyzed by a highly sensitive ion chromatography technique 
(using a ICS2000 system for anions and Dionex 500 system for cations)\cite{Motizuki2017}.
See Motizuki et al. (2017)\cite{Motizuki2017} for details of the analysis procedures.
Here, we report NO$_3^-$ concentrations and periodicities in the DF01 ice core for a $\sim$300-year period (1610--1904; corresponding core depths are 23.0--7.7 m).  
In this core segment, the precision of the NO$_3^-$ concentration measurements was reanalyzed in detail and found to be within 0.14--0.48 $\mu$g L$^{-1}$ at 10 $\mu$g L$^{-1}$, signified as the maximum, assuming that the errors deduced from each chromatogram are independent and adopting the law of error propagation. 
  
The firn top at 7.7 m depth in the DF01 core corresponds to the year 1904, whereas the ice at 85.5 m depth dates back to more than 2,000 years ago. \cite{Motizuki2014} 
The DF01 ice core was cut into shorter lengths of 50 cm at Dome Fuji, and then transported to the National Institute of Polar Research (NIPR). 
At NIPR, the 50-cm segments were further subdivided as follows: 
Those from depths shallower than 20 m were cut into 5-cm pieces, those from 20--50 m into 4-cm pieces, those from 50--75 m into 3-cm pieces, and those from deeper than 75 m into 2.5-cm pieces. 
Depending on the depth, the temporal resolution of the samples ranged from 0.7 to 1.0 year and was approximately 0.9 year on average. \cite{Motizuki2014}.

The chronology of the studied section of the DF01 ice core was established by synchronizing volcanic eruption signals in the DF01 ice core with corresponding signals in a reference core (the 100-m-deep B32 ice core collected from a site close to 
Kohnen station; Fig.~\ref{fig:Antarctica}) which was dated by counting annual layers.\cite{Motizuki2014}
Two time scales were established for DF01, DFS1 and DFS2, where DFS stands for ``Dome Fuji Shallow''. 
The DFS1 time scale, covering the period from CE 187 to 1904, was synchronized with the B32 ice core time scale by matching 31 volcanic eruption peaks in the non-sea-salt sulfate ion concentrations in the depth profiles of the two ice cores. 
The DFS2 time scale, covering the period from CE 1 to 1904, was based in part on published data\cite{Ruth2007} 
and also on four volcanic dates obtained from samples from the upper part of the 1,000-m-deep EPICA DML ice core drilled at the Kohnen station. 
The accumulation rates between neighboring volcanic eruptions, or time markers, were then assumed to be constant.
The dating error of the DF01 ice core is thus made up of the absolute error of the referenced time marker plus the interpolation error which depends on the temporal distance from the nearest time marker.\cite{Motizuki2014}
In the present study, we confined ourselves to the $\sim$300-year period from 1610 to 1904. 
The dating of this period was both robust and rather precise because it was based on many well-established volcanic time markers with absolute errors of 1--3 years.\cite{Motizuki2014} 
For this paper, we used the DFS2 time scale, but it should be noted that the DFS1 and DFS2 time scales for this period are identical.

\section{Solar signatures in the DF01 core} \label{sec:solarsig}

In this section, our results of time series analyses of NO$_3^-$ variations are presented and discussed.
 
\subsection{Time series of nitrate ion variations}\label{subsec:TS}
 
\begin{figure*}[ht!]
\begin{center}
\includegraphics[clip,width=17.5cm]{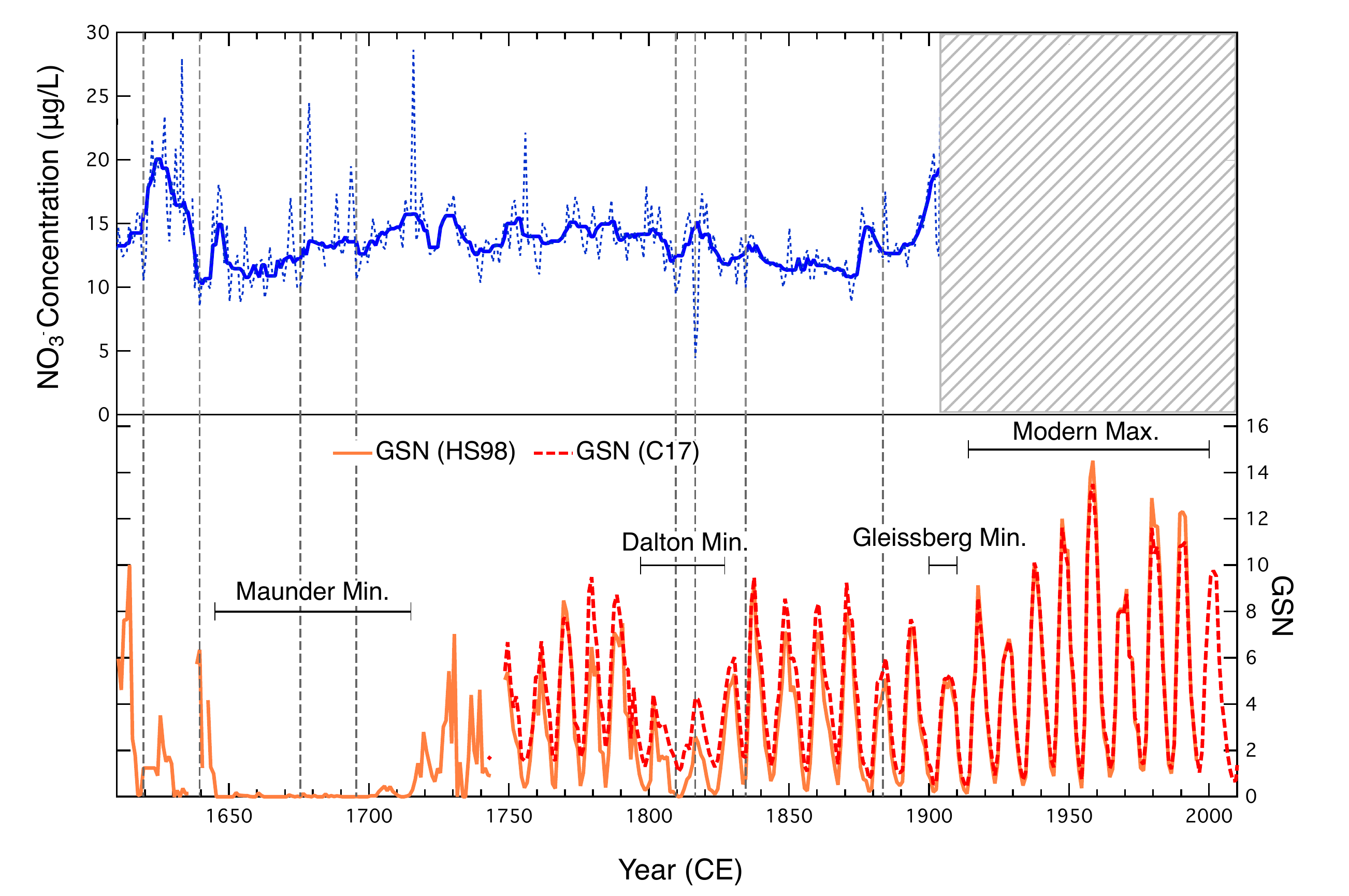}
\caption{
(Top) Time series of annually resolved NO$_3^-$ concentrations during 1610--1904 in the DF01 ice core. 
Raw data (a dashed blue line) and those after application of 7-point moving median smoothing (a blue line). 
(Bottom) Annual group sunspot numbers (GSN) proposed by Hoyt and Schatten (1998) (HS98) and Chatzistergos et al. (2017) (C17) 
are shown by a solid orange and a dashed red line, respectively. 
The vertical dashed gray lines indicate the positions of time markers used to date the DF01 ice core. 
These time markers represent the signals of the volcanic eruptions used to determine the DFS2 time scale.
The NO$_3^-$ concentrations (raw data) constitute negative spikes at the vertical gray lines because nitrate, weakly acidic, was displaced by coexsisting with sulfate, strongly acidic, originated from volcanic eruptions.
Durations of the grand solar minima, Maunder minimum and Dalton minimum, are indicated by bars.
\label{fig:NO3-GSN}}
\end{center}
\end{figure*}

The time series of NO$_3^-$ variations (raw data) in the DF01 ice core from 1610 to 1904, based on the DFS2 chronology,\cite{Motizuki2014} is depicted in Fig.~\ref{fig:NO3-GSN}. 
The raw data show meteorological noise (seen as positive spikes) on an interannual scale, as Traversi et al. (2012)\cite{Traversi2012} mentioned with regard to the Talos Dome core. 
The raw data may contain some meaningful spike structures; these analytic results will be reported separately. 
 ``Negative'' spikes, also seen in our raw data, correspond with the positions of time markers (vertical gray dashed lines in Fig.~\ref{fig:NO3-GSN}). 
These time markers represent the signals of the volcanic eruptions used to determine the DFS2 time scale. 
Negative spikes occurred where nitrate was displaced by sulfate that originated from volcanic eruptions, 
as mentioned in Sect.~\ref{sec:intro}.

To determine the baseline variation in the raw NO$_3^-$ concentrations and investigate NO$_3^-$ concentration modulations possibly embedded in our time-series data, we applied running median filters. 
Using the median instead of the mean minimized the risk that outliers stemming from other physical sources (Sect.~\ref{sec:intro}, mostly event-type) might have skewed the result.
We found that the time series of NO$_3^-$ variations obtained after applying a running 7-point (corresponding to 6-year) median filter to the raw NO$_3^-$ concentration time series  (Fig.~\ref{fig:NO3-GSN}) was appropriate since it was not affected by positive or negative spikes during the period studied. 
Note that the measurement imprecision (within 0.14--0.48 $\mu$g L$^{-1}$ at concentrations of 10 $\mu$g L$^{-1}$; see Sect.~\ref{sec:DF01}) 
was very small compared with the magnitude of the variations in the median-filtered NO$_3^-$ concentration time series (Fig.~\ref{fig:NO3-GSN}), which we therefore regard as baseline variations.
 A direct eye-inspection of our baseline variations in Fig.~\ref{fig:NO3-GSN} enables us to observe oscillations of $\sim$20 years.

In Fig.~\ref{fig:NO3-GSN}, annual group sunspot numbers (GSN) proposed by Hoyt and Schatten (1998)\cite{HoytSchatten1998} 
and Chatzistergos et al. (2017)\cite{Chatzistergos2017}
 (HS98 and C17, respectively) are also shown. 
These GSN profiles together cover the period from 1610 to 2010. 
The C17 GSN series calibrates the results from 314 observers since 1739 with a non-linear non-parametric method and may be one of the best current estimates.\cite{Munoz-Jaramillo2019}

\subsection{Periodicity in nitrate ion variations}\label{subsec:periodicity}
 
\begin{figure*}[ht!]
\begin{center}
\includegraphics[clip,width=16.0cm]{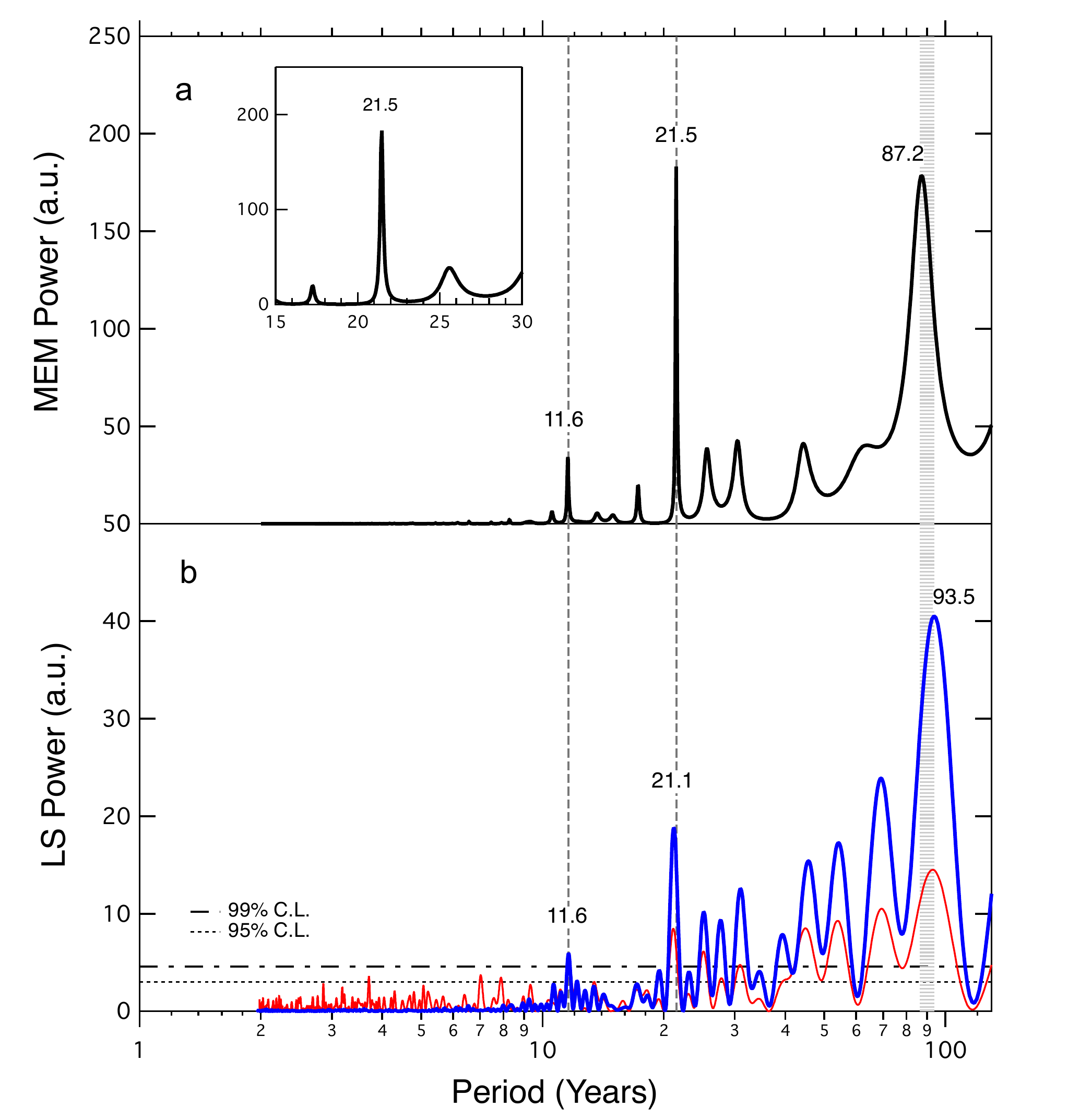}
\caption{
Power spectra for the baseline NO$_3^-$ concentration time series for 1610--1904 obtained by (a) the Maximum Entropy Method (MEM) for the 7-point median series
with the inset to show the 21.5-year peak signal, 
and (b) the Lomb-Scargle (LS) method (shown with confidence levels; C.L.). In (b), the red line shows the result obtained for the raw data, and the blue line shows that obtained for the 7-point median series.
\label{fig:MEM_LS_full}}
\end{center}
\end{figure*}

The Maximum Entropy Method (MEM)
\cite{Wu1997}\cite{Hino2010}
was used to detect periodicities in the median-filtered, baseline NO$_3^-$ concentration profile (Fig.~\ref{fig:MEM_LS_full}a). 
This method can generate high-resolution power spectra for a short, evenly spaced time series. 
We also applied the Lomb-Scargle (LS) method
\cite{Lomb1976}$^{-}$\cite{HorneBaliunas1986}
to our raw and median-filtered data (Fig.~\ref{fig:MEM_LS_full}b). 
The LS method is able to generate power spectra of adequate resolution even when applied to unevenly spaced time-series data or 
even if a portion of the data series is masked.
The application to our baseline data of these two distinct statistical methods generated well-matched peaks at 11.6, 21, and around 90 years (Fig.~\ref{fig:MEM_LS_full}). 
We confirmed that periodicities with these peaks also existed in the raw data (Fig.~\ref{fig:MEM_LS_full}b).
The power spectra of periodicities shorter than 10 years in the raw data are evident, as expected, because of the meteorological noise (Fig.~\ref{fig:NO3-GSN}). 
 
The 99\% and 95\% confidence levels (Fig.~\ref{fig:MEM_LS_full}b) were calculated as the probability that the height of an LS power of a given periodicity exceeds by 1\% and 5\%, respectively, chance variations due to random noise, if the random noise follows a normal distribution. 
(Note that equivalent confidence levels cannot be derived for MEM.) 
The peak values of the detected periodicities of around 11, 22, and 90 years are apparently higher than the 99\% confidence level (Fig.~\ref{fig:MEM_LS_full}b). 
These statistically significant NO$_3^-$ concentration periodicities of $\sim$11, $\sim$22, and $\sim$90 years are almost certainly related to the well-known 11-year Schwabe, 22-year Hale, and $\sim$80--90-year Gleissberg solar cycles.
\cite{Usoskin2017}$^{-}$\cite{PeristykhDamon2003} 
The simultaneous detection of the three known shortest solar periodicities in the NO$_3^-$ concentrations in an ice core has not been reported previously.
 
\subsection{Prominent 22-year periodicity and its band-pass filtering}\label{subsec:22-yearPeriodicity}

We will consider first about the prominent emergence of the 22-year signal obtained in Fig.~\ref{fig:MEM_LS_full}, because this is somewhat different from the understanding of the activity of the sun on the 11-year and 22-year cycles.
The 22-year solar periodicity is associated with the reversals of the magnetic dipole field polarity of the sun in every 11 years. 
The 22-year periodicity affects GCRs penetrating into the earth: 
When the solar activity is high, the GCR intensity reaching Earth weakens because the intense solar magnetic field in space prevents the GCRs from entering the internal space surrounding the Earth.   
It is well known that the power of the 22-year periodicity of the intrinsic solar activity is considerably smaller than that of the 11-year periodicity.

As mentioned in Sect.~\ref{sec:intro}, Watanabe et al. (1999)\cite{Watanabe1999}, using the MEM method, found an 11-year periodicity in their NO$_3^-$ concentration time series, but not a 22-year periodicity (their fig.~4), in a short firn core covering the years 1920--1980, a period overlapping the modern grand maximum (Fig.~\ref{fig:NO3-GSN}), obtained from the S25 site (Fig.~\ref{fig:Antarctica}). 
They confirmed that the S25 firn core was not affected by the photolysis in the surface snow because of the high snow precipitation rate.
In addition, by applying a 9--13-year bandpass filter to their NO$_3^-$ concentration data,
they found that the filtered 11-year NO$_3^-$ modulation was in phase with the filtered 11-year cycle in sunspot numbers (their fig.~5).
Taking into account the photochemical reactions occurring in the stratosphere, we can reasonably consider that when solar activity is high, the rate of production of NO$_{\rm y}$ should also be high, and vice versa (Fig.~\ref{fig:reaction}).
These give us a hint that the dominant NO$_{\rm y}$ production mechanism for the S25 core site would be caused by N$_2$O oxidation in the stratosphere (Fig.~\ref{fig:reaction}), not by GCRs, for which the bandpass result should have had inverse phases. 
   
We hypothesized that the intense $\sim$22-year signature in our DF01 core could be attributed to the photolysis in the surface snow that occurs at Dome Fuji but not at S25, and applied an 18--30 year band pass filter to both the baseline NO$_3^-$ concentration profile and to the HS98 and C17 sunspot number profiles (see Fig.~\ref{fig:NO3-GSN}) in order to investigate their ``22-year'' oscillations.
The bandwidth was determined by an observation of the 22-year peak in our NO$_3^-$ time series (see the inset in Fig.~\ref{fig:MEM_LS_full}a) as well as preceding work. 
The result is shown in Fig.~\ref{fig:filtered_NO3-GSN}a.
We also confirmed that applying a shorter bandpass range did not change the essence of the result.
Because we assumed that the ``22-year'' periodicity was embedded by the photolysis in the surface snow at Dome Fuji,
the NO$_3^-$ axis in Fig.~\ref{fig:filtered_NO3-GSN}a was reversed to see the inverse relationship between filtered NO$_3^-$ and GSN time series easier.
 
First, we point out in Fig.~\ref{fig:filtered_NO3-GSN}a that the amplitude of each ``22-year'' NO$_3^-$ oscillation is distinctly larger than the referenced maximum measurement error, mentioned in Sect.\ref{sec:DF01}:
Throughout the studied period, the ``22-year'' oscillations were stably identified.
We observe that the ``22-year'' oscillations were eminent during both the Maunder Minimum and the Dalton Minimum.
This means that there is a possibility that these 22-year NO$_3^-$ modulations could be used as a relatively reliable time measure for dating the deep Dome Fuji cores in the future.

Second, we see in Fig.~\ref{fig:filtered_NO3-GSN}a that 
the ``22-year'' filtered NO$_3^-$ oscillations are mostly in inverse phases with the  ``22-year'' filtered GSN oscillations throughout the total period, except phases in $\sim$1720--1780 and $\sim$1875, just after the Maunder Minimum and the Dalton Minimum. 
The photolysis in the surface snow is influenced by local variations in the UV intensity reaching the ground, air temperatures, and so forth.\cite{Traversi2017}.
The anticorrelation found in Fig.~\ref{fig:filtered_NO3-GSN}a might then indicate that the $\sim$22-year solar periodicity was superposed on the NO$_3^-$ concentrations in the DF01 core by the local loss of NO$_3^-$ through solar UV photolysis.
As mentioned in Sect.~\ref{sec:intro}, the photolysis process has been intensively studied,
\cite{Frey2009}$^{-}$\cite{Akers2022}
but much remains to be understood;
in addition, Fig.~\ref{fig:filtered_NO3-GSN}a requires further new data analyses to be comprehensively explained.

\subsection{``11-year'' band-pass filtering and cyclicity of solar activity in Maunder Minimum}\label{subsec:11-yearBPF}

Although the average duration of the 11-year solar cycle is 11 years, the duration of individual cycles varies between 9 years (e.g., solar cycles 2, 3, 8, and 22) and 14 years (13.6 years, solar cycle 4), according to the smoothed monthly sunspot number time series from the Solar Influences Data analysis Center (SIDC) (https://www.sidc.be/silso/cyclesminmax).
Here we applied an 8--16 year bandpass filter to the moving median-filtered NO$_3^-$ concentration profile and to the HS98 and C17 sunspot number profiles (see Fig.~\ref{fig:NO3-GSN}) as in the preceding work.\cite{Berggren2009}
The result is depicted in Fig.~\ref{fig:filtered_NO3-GSN}b. 
   
We see in Fig.~\ref{fig:filtered_NO3-GSN}b that the amplitude of each ``11-year'' NO$_3^-$ oscillation is again predominantly or marginally larger than the referenced maximum measurement error.
We also recognize that the ``11-year'' oscillations during both the Maunder Minimum and the Dalton Minimum exceed the maximum measurement errors,
while during the Maunder Minimum with almost no sunspots (see Fig.~\ref{fig:NO3-GSN}), no 11-year oscillations are discernible in our bandpass-filtered result for the HS98 time series of group sunspot numbers\cite{HoytSchatten1998}.
This may indicate the existence of the cyclic behavior of solar dynamo even during the grand Maunder Minium, as suggested by preceding $^{10}$Be and $^{14}$C studies.
\cite{Berggren2009}\cite{Beer1998}\cite{Miyahara2004}
  
Next, our  ``11-year'' bandpass-filtered NO$_3^-$ concentration profile and the sunspot number profiles exhibit both inphase and reverse phase correlations.
They are:
1) in phase around the years 1620 and 1800--1820; the time markers for 1619 (unknown eruption) and 1816 and 1809 (Tambora and pre-Tambora eruptions) make this view relatively certain; 
2) in phase during the years around 1720--1760, but we need to be cautious because the dating of DF01 samples covering these years is less precise as the two closest time marker positions are distant, at 1696 and 1810; thus, the absolute error plus the interpolation error of the DF01 sample dates for around 1750 can be $\sim$5 years;\cite{Motizuki2014} 
3) in reverse phase during the years 1860--1910, the time marker for 1883 (Kratakau, Indonesia) making this observation reliable.  

Regarding the present result of Fig.~\ref{fig:filtered_NO3-GSN}, the stratospheric NO$_{\rm y}$ production by the N$_2$O oxidation and the loss of NO$_3^-$ by the photolysis in the surface snow might, together, affect the NO$_3^-$ concentration profile at the Dome Fuji site, but to draw conclusions will require further careful investigation.

\section{Concluding remarks} \label{summary}
 
The DF01 ice core drilled at Dome Fuji station in East Antarctica has been shown to demonstrate substantial sensitivity to variations in atmospheric NO$_{\rm y}$  in the Antarctic polar stratosphere.
This was possible mainly because of: 1) a local feature of the Dome Fuji ice core, which appears to retain 
more of the NO$_{\rm y}$ stratospheric budget than some ice cores obtained elsewhere;  
2) continuous, annually-resolved precision NO$_3^-$ measurements and error analyses; and 3) reliable dating of the ice core.

\begin{figure*}[htb!]
\begin{center}
\includegraphics[clip,width=17.5cm]{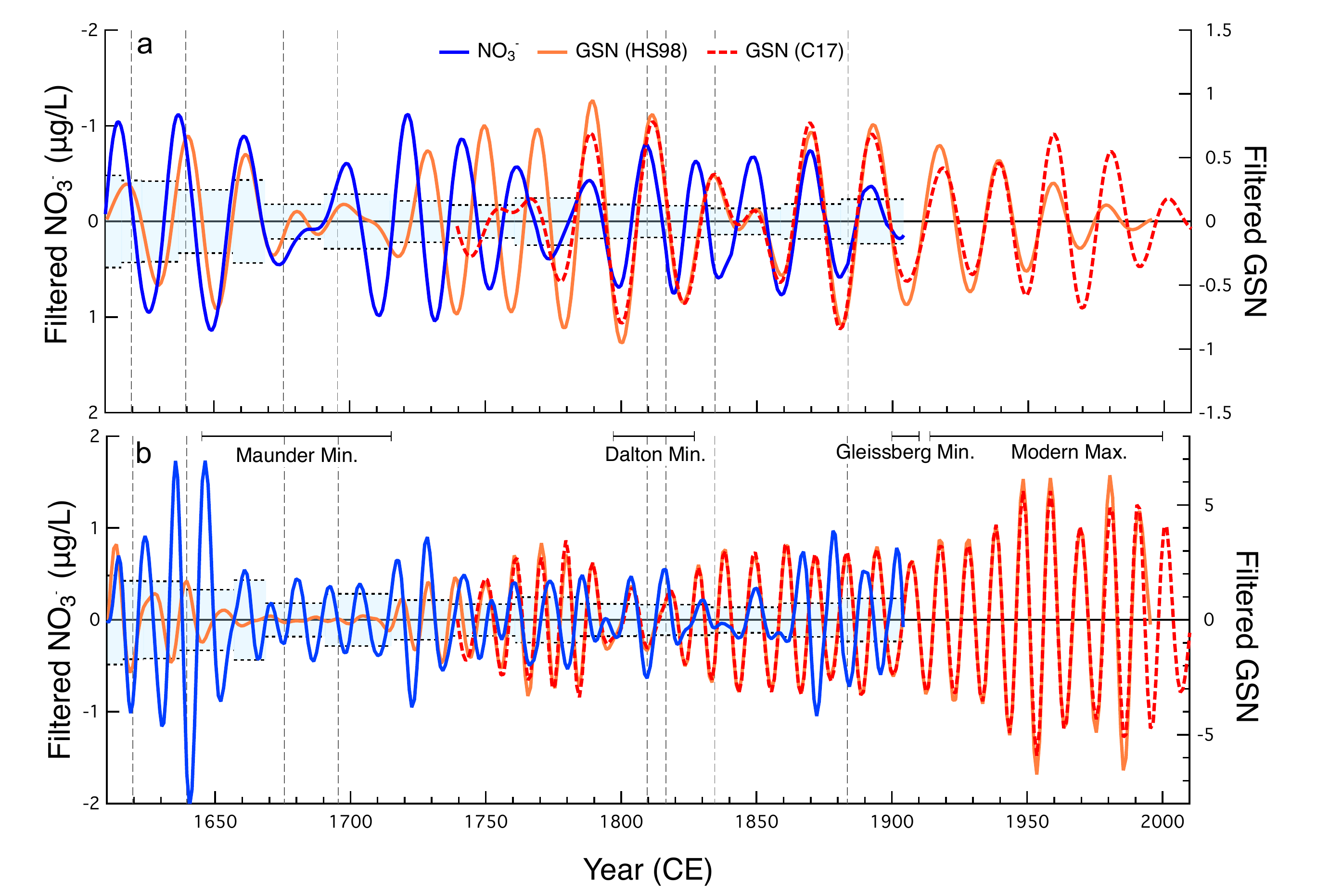}
\caption{
Bandpass-filtered variations in the baseline NO$_3^-$ concentration (blue) and group sunspot numbers (solid orange and dashed red lines) from 1610 to 2010. 
Vertical dashed gray lines indicate the positions of the time markers in the DF01 ice core.
a) Bandpass filtered results  with the band from  18 to 30 years show ``22-year'' modulations. 
b) Those with the band from 8 to 16 years show ``11-year'' modulations. 
Both results, a) and b), were obtained using a Butterworth filter.
The light blue highlighted region indicates the estimated maximum error in our NO$_3^-$ measurements.
Note that the range of the left axes of a) and b) are the same but reversed in a).
Durations of the grand solar minima and modern solar maximum are indicated by bars in the center (see also Fig.~\ref{fig:NO3-GSN}).
\label{fig:filtered_NO3-GSN}}
\end{center}
\end{figure*}

Our time series analysis results for  NO$_3^-$ concentrations in the DF01 ice core segment covering the period from 1610 to 1904 revealed $\sim$11-year, $\sim$22-year, and $\sim$90-year periodicities, most probably corresponding to the three well-known decadal to multi-decadal solar cycles.
These results represent the first finding of these three shorter cycles in the NO$_3^-$ concentrations in ice cores at the same time. 
We propose that the NO$_3^-$ concentration profile in the Dome Fuji ice core be used as a new proxy for these shorter solar cycles.
The 22-year and 11-year modulations were seen both in inverse phases and in phases with respect to the sunspot number modulations.
Among the modulations, it was suggested that the 22-year periodicity might have been superposed by the photolysis that occurs in the surface snow; 
we hypothesized this since the 22-year modulations were intense and mostly in inverse phases with respect to the sunspot number modulations, except just after the Maunder Minimum and the Dalton Minimum.
Regarding the present result in Fig.~\ref{fig:filtered_NO3-GSN}, both the stratospheric NO$_{\rm y}$ production by N$_2$O oxidation and the local effect of the loss by the photolysis in the surface snow could affect the NO$_3^-$ concentration profile at the Dome Fuji site, but to draw conclusions will require further careful investigation. 
 
Finally, 
we have pointed out the possibility of using the relatively stable ``22-year'' NO$_3^-$ modulations as a time measure to date deep ice cores, at least within the core from the same drilling hole (DF2; its bottom reached $\sim$720,000 years ago; Sect.\ref{sec:DF01}). 
This could be possible even when an annual layer thickness becomes sub-centimeter in deeper depths; such a high-resolution sampling has already been a target, using our new laser-melting method for ice cores (Motizuki et al., submitted for publication). 
Finding these decadal to multi-decadal cycles in other Dome Fuji shallow and deep ice cores would be an interesting challenge, and disentangling the mechanism underlying the results reported here is very important to an understanding of the profiles of NO$_3^-$ as it is a major chemical components observed in ice cores.
The use of annually resolved NO$_3^-$ concentrations in Dome Fuji ice cores as a new potential proxy for solar activity should increase the value of such measurements. 
 
\bigskip
 
\begin{center}
\textbf{Acknowledgments}
\end{center}

We are deeply grateful to K. Makishima for enlightening discussions as well as for his support for our research team. 
We are grateful to M. Igarashi for his early analyses for this project. 
We thank Y. Fujii for his crucial role in initiating this astronomical collaboration as well as for his continuous encouragement, 
and we acknowledge the help of K. Kamiyama and T. Ohata in starting this collaboration. 
We also thank K. Suzuki, Y. Iizuka, H. Akiyoshi, T. Yokoyama, and A. Asai for valuable comments. 
We acknowledge 
H. Sakurai and Y. Yano for discussions on the measurement precision and Y. V. Sahoo for repeating the error check.
We are indebted to M. Kitagawa for help in measuring the ion concentrations in this work. 
Finally, we are grateful to the members of the 42nd 
Japanese Antarctic Research Expedition and the Dome Fuji drilling team for supplying the DF01 ice core for this work and to all of the members who participated in sampling the DF01 core at NIPR. 
This research was supported in part by a JSPS Grant-in-Aid for Scientific Research (A) (Grant Number 22244015), the NEXT Program of CSTI (Grant Number GR098), and 
research funds provided by RIKEN Nishina Center.

\clearpage

\end{document}